Reduction of a Single Layer Graphene Oxide Film on Pt(111)

Seiji Obata[1], Hiroshige Tanaka[1], and Koichiro Saiki[1,2]

[1]Department of Chemistry, The University of Tokyo, Kashiwanoha 5-1-5, Kashiwa, Chiba 277-8561, Japan

[2]Department of Complexity Science and Engineering, The University of Tokyo, Kashiwanoha 5-1-5, Kashiwa, Chiba 277-8561, Japan

Graphene oxide (GO) is a very attractive material from which graphene of mass quantity could be fabricated. To restore the outstanding properties of graphene, however, complete reduction of GO is necessary, which no one has ever achieved before. We examined annealing of GO which was placed on Si(100) and Pt(111) and observed the surface atomic structure by scanning tunneling microscopy. A honeycomb lattice with long range order appeared for GO on Pt(111), but not for GO on Si(100). Reduction of GO together with restoration of the graphene lattice might be realized by the catalytic property of Pt, which opens a new way to synthesize graphene.

Graphene is a single atomic layer of carbon arranged in a hexagonal (honeycomb) lattice[1]. Its fascinating transport properties, such as an unusual quantum Hall effect[2, 3] and a field effect mobility as high as 200000 cm$^2$/ (V·s)[4], have made research on graphene widespread explosively. By considering technological applications, the development of high throughput methods of preparing graphene is strongly required. Although mechanical exfoliation of graphite, with which Novoselov et al. first succeeded in preparing graphene[1], gives the highest quality specimen, this method is not suitable for use in the electronics industry. Techniques other than the exfoliation technique, such as the graphitization of SiC[5], chemical vapor deposition (CVD) on metal substrates[6] and a chemical preparation process via graphene oxide (GO)[7, 8], have been devised so far. Although the chemical preparation method seems suitable for mass production, reduction from GO to graphene causes a severe problem to be solved. The typical reduction methods that have been employed so far are heating and chemical reaction with a reducing agent, but they have been proved quite insufficient. For example, the field effect mobility of reduced GO (rGO) is typically 10-400 cm$^2$/(V·s)[9,10], which is smaller by two or three orders of magnitude than that of exfoliated graphene. Raman spectra and X-ray photoelectron spectra (XPS) of rGO indicate the existence of lattice disorder and the remaining hydroxyl group, respectively[11), 12]. Under these circumstances, a novel reduction method for GO is urgently required.

Here, we report a new technique for producing high quality graphene from GO. We annealed GO flakes placed on Pt(111) and Si(100) in vacuum at around 1300 K and observed the GO surface by scanning tunneling microscopy (STM). A honeycomb lattice was clearly observed and moiré structures indicating a long range order were observed for GO on Pt(111), while an atomic order was observed only partially for GO on Si(100). The result of photoemission measurement showed no trace of oxygen for rGO on Pt(111), while the remaining oxygen was observed for rGO on Si(100). It can be concluded that the annealing of GO on Pt(111) produces high quality graphene sheets, which might be realized with the help of the catalytic property of a Pt(111) surface.

GO was synthesized by the modified Hummers method[13), 14)]. This method oxidized graphite powder (mean size 20 μm, SEC carbon) using $KMnO_4$, $H_2SO_4$ and $NaNO_3$. After oxidation, GO was washed with $H_2O_2$ and $H_2SO_4$ solutions followed by repeated rinsing with pure water. The GO film was prepared by dipping the substrate into the solution. The water solution of GO is, however, not suitable for deposition probably because of its large surface tension. Addition of MeOH to the water solution makes it easy to deposit GO on various substrates. We used two kinds of substrates, Pt(111) and highly doped n-type Si(100). Pt(111) was mechanically polished and washed with acetone, and Si(100) was cleaned by UV/ozone treatment. The thickness or the coverage of the GO film was controlled by changing the dipping time or the density of GO solution.

First, we examined annealing of GO/Si(100) in vacuum. Figure 1(a) shows a typical STM image of GO on highly n-doped Si(100), which was annealed in vacuum at 1050 K for 30 min. Each bright spot (some bright spots are shown by dots) represents one carbon atom, which we can judge from the interval between spots. Although each carbon atom can be observed clearly, a periodic pattern or long range order beyond a few nanometers was not observed. There is neither a honeycomb lattice nor a triangle lattice, which is characteristic of a single layer graphene or highly oriented pyrolytic graphite (HOPG), respectively. Annealing to 1300 K did not improve the long range order of GO and further annealing was likely to cause chemical reaction with the Si surface. We observed the STM images from six different regions, all of which were similar to that shown in Fig. 1(a).

The electrical conductivity of GO reduced by various processes indicates that hydrazine treatment before thermal annealing is effective for improving the field effect mobility[12)]. Figure 1(b) shows a typical STM image of GO on Si(100) after the hydrazine treatment and the subsequent thermal annealing at 1050 K. The region in which a honeycomb lattice is observed as shown in the inset can be found in many places, while a periodic pattern is limited within a few nanometers. This means that the graphene lattice is restored partially, but the regions with a honeycomb lattice are rather isolated with poor connectivity. The residue of the disordered structure indicated by circles in Fig. 1(b) causes a much lower field effect mobility of rGO than the exfoliated graphene.

The above STM observations have revealed that only limited reduction was achieved by the reducing agent and thermal annealing. An alternative method should be devised to enhance the reduction of GO further. One promising way is to use the catalytic property of a metal surface to reduce GO. This can be expected from the fact that graphene grows through the CVD process on the surfaces of metals such as Pt(111)[15,16,17], Ni(111)[18], Ir(111)[19], Ru(0001)[20], Cu(111)[21], Cu foil[6], Ni foil[22], and so on. In our previous work, we investigated growth of graphene from benzene molecules on Pt(111)[16], and observed the atomic structure together with moiré patterns[17]. In the present work, we annealed GO/Pt(111) in ultrahigh vacuum (UHV). Before the reduction experiment of GO, a Pt(111) substrate dipped in MeOH solution without GO was annealed at 1300 K in a similar UHV condition. The XPS of Pt(111) showed neither carbon nor oxygen signals and the STM image showed a clean terrace structure with atomic height steps. Thus the clean Pt surface is considered to be used in the following experiment.

Figure 2(a) shows a typical STM image of GO on Pt(111), which was annealed at 1300 K. There are steps just like on a clean Pt(111) surface. The step height is 0.22-0.24 nm which is in close agreement with that on Pt(111). Between the steps, a flat terrace can be observed. A magnified image of the white square (3.8 × 3.8 nm$^2$) in Fig. 2(a) is shown in Fig. 2(b). A honeycomb lattice which corresponds to a single layer of graphene is clearly observed. Any region in Fig. 2(a) shows a similar atomic image to Fig. 2(b) and the step height is always equal to that on Pt(111). Therefore, it can be thought that the graphene sheet reduced from GO covers uniformly a Pt(111) surface over a wide range.

The graphene grown on Pt(111) by pyrolysis of hydrocarbons shows a moiré structure.[17] The present rGO specimen also shows a moiré structure, a typical example of which is shown in Fig. 3(a). A magnified image of Fig. 3(a) is shown in Fig. 3(b), in which a honeycomb lattice appears with atomic resolution. In the lower part, a clear superstructure is seen, while in the upper part, it is not apparently seen. However, a closer observation of the STM image together with Fourier analysis [see insets in Fig. 3(b)] indicates the existence of a different superstructure in the upper region. A precise analysis reveals $\sqrt{31} \times \sqrt{31}$ R-8.9° and $7 \times 7$ R-21.8° structures in the upper and lower regions, respectively. These moiré patterns can be reproduced by superposing a graphene lattice on a Pt(111) substrate. The model of $7 \times 7$ R-21.8° structure is indicated in Fig. 3(c). This fact supports formation of a single layer graphene from GO on Pt(111).

The boundary between the different moiré patterns is ascribed to the grain boundary of graphene, because the height profile across such boundaries shows no difference in height. Another type of boundary relates with a step, a typical example is seen in Fig. 3(a), running vertically from top to bottom. The height difference is just the same as the step height on Pt(111). It should be noted that the moiré patterns of the same symmetry are observed on both sides. This means that the orientation of graphene is the same in both upper [left-hand side of Fig. 3(a)] and lower [right-hand side] terraces. The continuity of a moiré pattern over the step is observed in many positions at various steps as shown in Fig. 2(a). These results clearly indicate that a single layer graphene sheet lies over the step like a carpet. It is likely that the GO placed over the step of Pt(111) was reduced at the original site. In some places, we can find some kind of defect as shown in Fig. 3(d). However, the array of atoms surrounding this region looks continuous, which can also be supported by the superposed moiré structure prevailing in the whole region. We think that some contaminant on Pt(111) causes this disorder, but the graphene sheet seems to cover it flexibly.

The appearance of the ordered lattice and the moiré structure indicate that annealing reduces GO on Pt(111) and the graphene lattice is considerably restored, while annealing of GO on Si(100) results in partial reduction. We measured the XPS spectra of rGO *in situ* both on Pt(111) and Si(100). Figure 4(a) shows C 1s and O 1s signals before and after annealing both on Pt(111) and Si(100). Before annealing, oxygen appeared both on GO/Pt(111) and GO/Si(100). Annealing causes disappearance of oxygen on GO/Pt(111), while oxygen remains on GO/Si(100). Careful inspection of the Si 2p region (not shown) indicates the absence of the Si-O bond; thus, most oxygen signals come from the oxygen atoms remaining in rGO. The C 1s peak consists of several components before annealing for example, C-C, C-OH, and C=O. After annealing, the C-C peak becomes dominant for rGO on both Pt(111) and Si(100). The high resolution spectrum of the C1s peak is shown in Fig. 4(b). For comparison, the peak intensity and energy are set at the same value. After annealing at 1300 K, the peak of rGO on Pt(111) becomes as narrow as that of the HOPG bulk specimen. In contrast, the peak of rGO on Si(100) is much wider than that of HOPG, mainly because a small amount of the C-OH component remains even after annealing. This phenomena also indicates that the catalytic property of Pt(111) enhances the reduction of GO and provides graphene of high quality.

In summary, we found a novel method for synthesizing a high quality graphene from GO, which is easily and reproducibly prepared from graphite powder at small cost. The annealing of GO on Pt(111) at around 1300 K removes oxygen completely from the GO layer, which could not be achieved by hydrazine and thermal annealing on Si(100). The atom resolved STM images show a long range order in the reduced GO monolayer sheet (more than $50 \times 50$ nm$^2$), extending over the Pt steps. The present method is more advantageous than the CVD process on the metal substrate, because in the latter case, the steps and remaining contaminant are likely to work as an obstacle for the extension of the graphene sheet.

**Acknowledgments**

This work was supported by a Grant-in-Aid for scientific research from the Ministry of Education, Culture, Sports, Science & Technology (MEXT) of Japan. One of the authors (S.O.) would like to acknowledge the Global Centers of Excellence (COE) Program "Chemistry Innovation through Cooperation of Science and Engineering", MEXT, Japan, for financial support.

References

1) K. S. Novoselov, A. K. Geim, S. V. Morozov, D. Jiang, Y. Zhang, S. V. Dubonos, I. V. Grigorieva, and A. A. Firsov: Science **306** (2004) 666.

2) K. S. Novoselov, A. K. Geim, S. V. Morozov, D. Jiang, M. I. Katsnelson, I. V. Grigorieva, S. V. Dubonos, and A. A. Firsov: Nature **438** (2005) 197.

3) Y. Zhang, Y. W. Tan, H. L. Stormer, and P. Kim: Nature **438** (2005) 201.

4) J. H. Chen, C. Jang, S. Xiao, M. Ishigami, and M. S. Fuhrer: Nat. Nanotechnol. **3** (2008) 206.

5) C. Berger, Z. Song, X. Li, X. Wu, N. Brown, C. Naud, D. Mayou, T. Li, J. Hass, A. N. Marchenkov, E. H. Conrad, P. N. First, and W. A. de Heer: Science **312** (2006) 1191.

6) X. Li, W. Cai, J. An, S. Kim, J. Nah, D. Yang, R. Piner, A. Velamakanni, I. Jung, E. Tutuc, S. K. Banerjee, L. Colombo, and R. S. Ruoff: Science **324** (2009) 1312.

7) V. C. Tung, M. J. Allen, Y. Yang, and R. B. Kaner: Nat. Nanotechnol. **9** (2008) 1.

8) G. Eda, G. Fanchini, and M. Chhowalla: Nat. Nanotechnol. **3** (2008) 270.

9) C. Y. Su, Y. Xu, W. Zhang, J. Zhao, X. Tang, C. H. Tsai, and L. J. Li: Chem. Mater. **21** (2009) 5674.

10) S. Wang, P. K. Ang, Z. Wang, A. L. L. Tang, J. T. L. Thong, and K. P. Loh: Nano Lett. **10** (2010) 92.

11) D. Yang, A. Velamakanni, G. Bozoklu, S. Park, M. Stoller, R. D. Piner, S. Stankovich, I. Jung, D. A. Field, C. A. Ventrice, Jr., and R. S. Ruoff: Carbon **47** (2009) 145.

12) S. Obata, H. Sato, H. Tanaka, and K. Saiki: in preparation for publication.

13) W. S. Hummers, Jr. and R. E. Offeman: J. Am. Chem. Soc. **80** (1958) 1339.

14) M. Hirata, T. Gotou, S. Horiuchi, M. Fujiwara, and M. Ohba: Carbon **42** (2004) 2929.

15) T. A. Land, T. Michely, R. J. Behm, J. E. Hemminger, and G. Comsa: Surf. Sci. **264** (1992) 261.

16) S. Entani, S. Ikeda, M. Kiguchi, K. Saiki, G. Yoshikawa, I. Nakai, H. Kondoh, and T. Ohta: Appl. Phys. Lett. **88** (2006) 153126.

17) M. Yamamoto, S. Obata, and K. Saiki: Surf. Interface Anal. **42** (2010) 1637.

18) R. Rosei, M. De Crescenzi, F. Sette, C. Quaresima, A. Savoia, and P. Perfetti: Phys. Rev. B **28** (1983) 1161.

19) J. Coraux, A. T. N¢Diaye, C. Busse, and T. Michely: Nano Lett. **8** (2008) 565.

20) A. L. V. de Parga, F. Calleja, B. Borca, M. C. G. Passeggi, Jr., J. J. Hinarejos, F. Guinea, and R. Miranda: Phys. Rev. Lett. **100** (2008) 056807.

21) L. Gao, J. R. Guest, and N. P. Guisinger: Nano Lett. **10** (2010) 3512.

22) Q. Yu, J. Lian, S. Siriponglert, H. Li, Y. P. Chen, and S. S. Pei: Appl. Phys. Lett. **93** (2008) 113103.

Figure captions

Figure 1.

(a) 9.5 × 9.5 nm$^2$ STM image of rGO after annealing at 1050 K for 30 min on Si(100). The image was taken with a sample bias voltage of $V_s$ = -1V and a tunnel current of $I_t$= 1 nA. The dots represent bright spots corresponding to carbon atoms.

(b) 6.5 × 6.5 nm$^2$ STM image of rGO after hydrazine and subsequent thermal annealing at 1050 K for 30 min on Si(100). The image was taken with a sample bias voltage of $V_s$ = -0.25 V and a tunnel current of $I_t$= 1 nA. The 1.7 × 1.7 nm$^2$ region surrounded by a rectangle is magnified in the inset. The hexagon has sides of 1.42Å.

Figure 2.

STM images of rGO on Pt(111) after 1300 K annealing for 1 h. (a) 100 × 100 nm$^2$ STM image taken with a sample bias voltage of $V_s$ = 0.05 V and a tunnel current of $I_t$= 3 nA. Each of A-A', B-B' and C-C' means a single crystalline domain lying over the Pt(111) steps. (b) 3.8 × 3.8 nm$^2$ STM image of a rectangular region in (a). White hexagons have sides of 1.42Å.

Figure 3.

Moiré structures of rGO on Pt(111) after 1300 K annealing for 1 h. All hexagons in the figure have sides of 1.42Å. (a) 17 ×17 nm$^2$ STM image taken with a sample bias voltage of $V_s$ = 0.01 V and a tunnel current of $I_t$= 5 nA. (b) 7.2 × 7.2 nm$^2$ STM image, corresponding to the rectangular region in (a), taken with a sample bias voltage of $V_s$ = 0.003 V and a tunnel current of $I_t$= 5 nA. Fourier transformation of the upper and lower regions are shown in the insets. Scale bars in the insets are 2 nm$^{-1}$. (c) The moiré structure (7 × 7 $R$-21.8°) is constructed by superposition of the graphene lattice (purple lattice) on the Pt(111) surface (green balls). (d) 8.4 × 8.4 nm$^2$ STM image taken with a sample bias voltage of $V_s$ = 0.1 V and a tunnel current of $I_t$ = 3 nA.

Figure 4.

(a) XPS spectra of C1s and O1s measured with a Mg Kα source before and after reduction of GO on Pt(111) and Si(100). Peaks around 520 eV originate from Pt 4p$_{3/2}$. (b) High resolution XPS spectra of C1s.

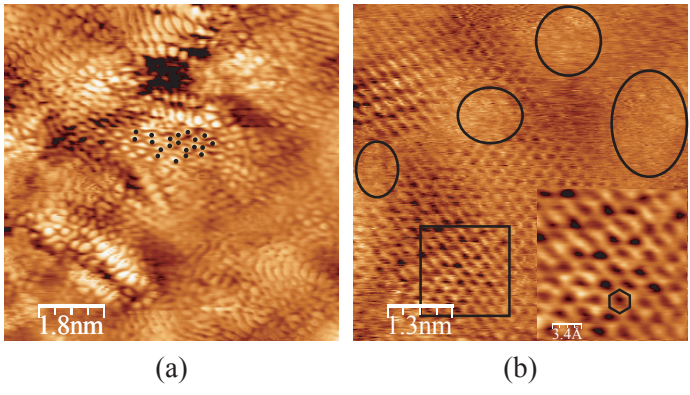

Fig 1

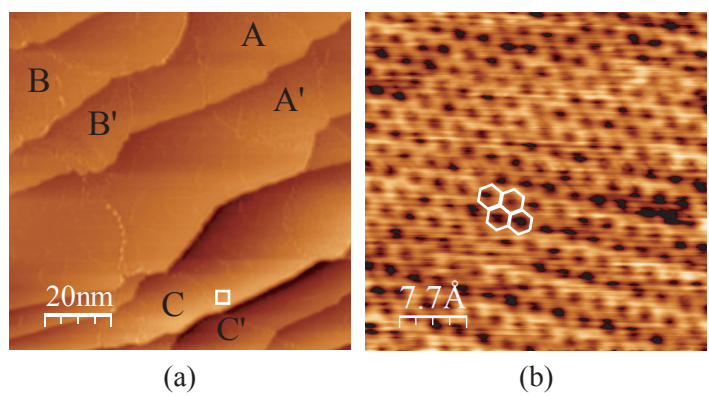

Fig 2

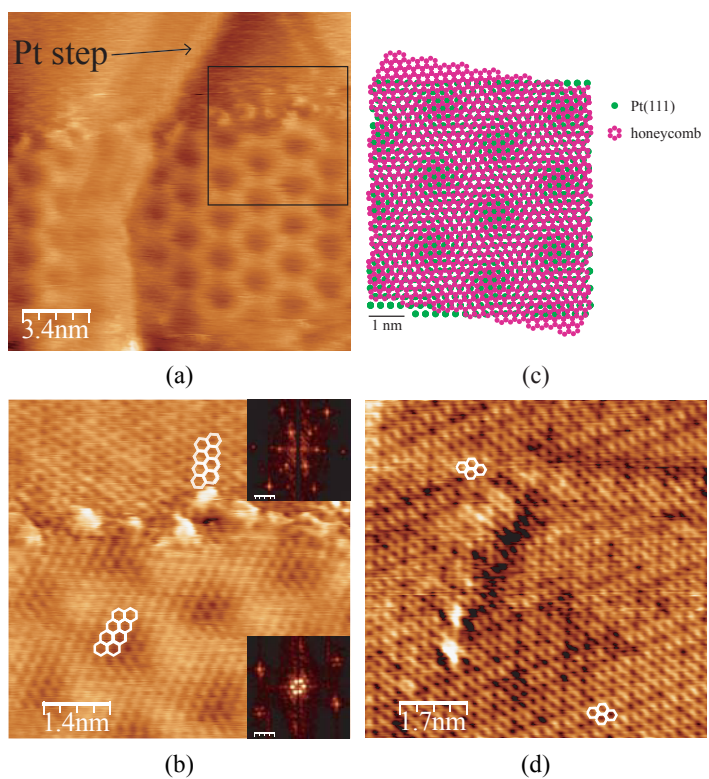

Fig 3

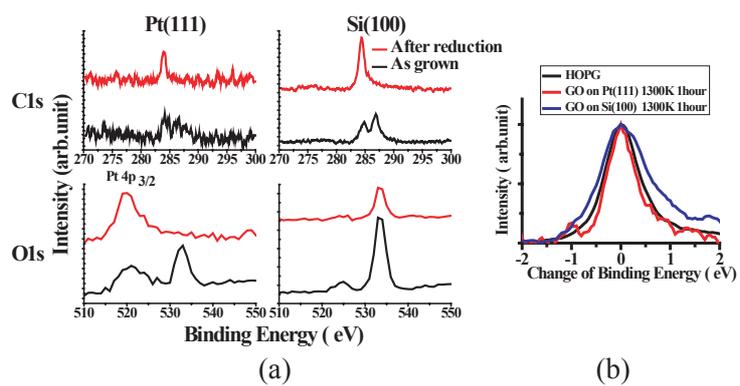

Fig 4